\documentstyle[amssymb,aps,multicol,epsf]{revtex}

\begin{document}
\draft

\title{
Pseudofractal Scale-free Web 
}

\author{
S.N. Dorogovtsev$^{1, 2, \ast
}$, A.V. Goltsev$^{2,\dagger}$, 
and J.F.F. Mendes$^{1, \ddagger}$
}

\address{
$^{1}$ Departamento de F\'\i sica and Centro de F\'\i sica do Porto, Faculdade 
de Ci\^encias, 
Universidade do Porto\\
Rua do Campo Alegre 687, 4169-007 Porto, Portugal\\
$^{2}$ A.F. Ioffe Physico-Technical Institute, 194021 St. Petersburg, Russia 
}

\maketitle
   
\begin{abstract} 
We find that scale-free random networks are excellently modeled 
by a deterministic graph. 
This graph has a discrete degree distribution (degree is the number of connections of a vertex) which is characterized by a power-law with exponent $\gamma=1+\ln3/\ln2$. 
Properties of this simple structure are surprisingly close to those of growing random scale-free networks with $\gamma$ in the most interesting region, between $2$ and $3$. 
We succeed to find exactly and numerically with high precision all main characteristics of the graph.
In particular, we obtain the exact shortest-path-length distribution. 
For the large network ($\ln N \gg 1$) the distribution tends to a Gaussian of width $\sim \sqrt{\ln N}$ centered at $\overline{\ell} \sim \ln N$. 
We show that the eigenvalue spectrum of the adjacency matrix of the graph  
has a power-law tail with exponent 
$2+\gamma$.   
\end{abstract}


\begin{multicols}{2}
\narrowtext


The essence of the modern situation in network science is the change-over 
from study of classical random graphs with Poisson degree distributions 
\cite{er59,er60} to exploration of complex networks with fat-tailed degree distributions \cite{ba99,s01,ab01a,dm01c,bck01,k01}. The prominent particular case of such nets are networks with power-law degree distributions (scale-free networks) \cite{ba99}. While growing, such nets actually self-organize into scale-free structures.   
These networks play a great role in Nature \cite{s01,ab01a,dm01c}. The Internet, the WWW, and many basic biological networks belong to this class. Fat tails of the degree distributions produce a number of intriguing effects \cite{ajb00d,ceah00a,cnsw00,ceah00b,pv01,dm01e}. 

Such networks are widespread, but very little is still known even about their basic properties \cite{krl00,dms00}. Most of real growing scale-free networks have $\gamma$ exponent of the degree distribution $P(k) \sim k^{-\gamma}$ in the range $(2,3)$, but this case turned to be the most difficult and unexplored. 
In particular, no exact results for the average shortest-path length $\overline{\ell}$ are known in this situation. 
The only known exact shortest-path length distributions were obtained for the simplest equilibrium networks \cite{dm00,k01a}. 
Notice that if $\gamma \leq 3$, standard estimates 
of $\overline{\ell}$ \cite{nsw00} are inapplicable to equilibrium networks with uncorrelated vertices. 
Correlations in growing networks are inevitable, and the results are even less encouraging. 
The generic property of these networks, which makes their analytical study so hard, is 
a  
complex structure of their adjacency matrices.    

Scale-free random networks naturally have a continuous degree distribution spectrum, but it has recently been shown that discrete degree distributions of some deterministic graphs also have a power-law decay \cite{br01}. 
The leading idea of the present work is the following. If it is so hard to get exact or precise results for scale-free random networks, especially, with $2 < \gamma < 3$, let us (i) construct a simple deterministic scale-free graph with such $\gamma$, 
(ii) obtain exact (analytical) and precise (numerical) answers for main structural and topological characteristics of the graph,  
(iii) compare its properties with known characteristics of corresponding scale-free growing random networks, and 
(iv) if these properties coincide, make new predictions for scale-free growing random networks. 
We actually model scale-free networks with $2 < \gamma < 3$ by the deterministic graph, whose structure can be described completely.  

Here we present results of this program. We succeed to find a number of exact characteristics of the scale-free deterministic graph, which are still unknown for random scale-free networks. The structural properties of deterministic and random scale-free growing networks proved to be surprisingly close to each other, so that our results can be reasonably applied to random growing nets.

\begin{figure}
\epsfxsize=85mm
\centerline{\epsffile{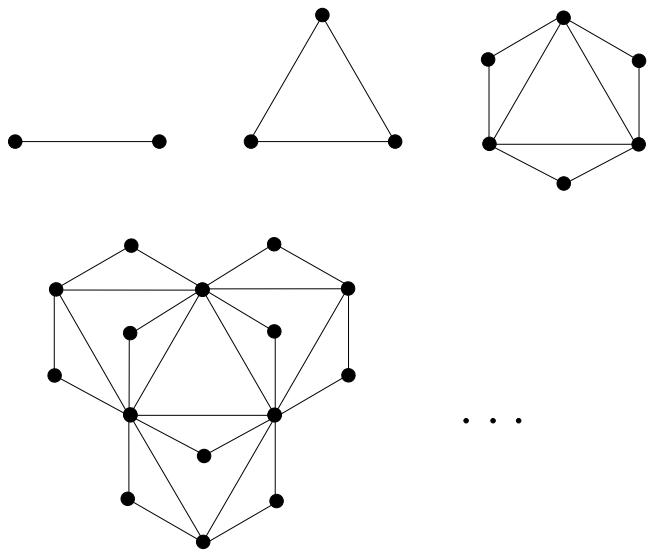}}
\caption{
Scheme of the growth of the scale-free pseudofractal graph. The growth starts from a single edge connecting two  vertices at $t=-1$. At each time step, every edge generates an additional vertex, which is attached to both 
end vertices of the edge. 
Notice that the graph at timestep $t+1$ can be made by connecting together the three $t$-graphs.    
}
\label{f1}
\end{figure}

{\em Pseudofractal graph}.---The growth starts from a single edge connecting two vertices at $t=-1$ (see Fig. \ref{f1}). At each time step, to every edge of the graph, a new vertex is added, which is attached to both the end vertices of the edge. Then, at $t=0$, we have a triangle of edges connecting a triple of vertices, at $t=1$, the graph consists of $6$ vertices connected by $9$ edges, and so on.  
The total number of vertices at ``time'' $t$ is $N_t=3(3^t+1)/2$, and the total number of edges is $L_t = 3^{t+1}$, so that the average degree is 
$\overline{k}_t = 2L_t/N_t = 4/(1+3^{-t})$. 

This simple rule   
produces a complex growing network which  
is certainly not a fractal \cite{remark}. Indeed, at any step, the entire graph can be set inside of a unit triangle. This means that the structure has no any fixed finite fractal dimension. On the other hand, one can depict the graph in another way, namely, like in Fig. \ref{f1} where the graph is surrounded by a long chain of edges.  
The length of this ``perimeter'' is $P_t = 3 \cdot 2^t$ edges, whence $N_t \sim L_t \sim P_t^{\ln3/\ln2}$. 
We can, however, fold the graph into a more compact structure with a different border, so that $\ln3/\ln2$ is not a fractal dimension of the structure but only some characteristic 
value. We failed to introduce a well defined spectrum of fractal dimensions (between $\ln3/\ln2$ and $\infty$), hence the network cannot be called a multifractal. 
Thus, this graph is not a fractal but only parody of it, and we call it, for brevity, {\em pseudofractal}. 
Notice that the graph contains numerous loops and hence is very far from tree-like.  

\begin{figure}
\epsfxsize=57mm
\centerline{\epsffile{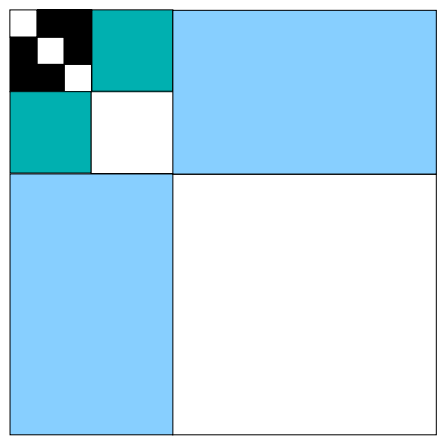}}
\caption{
Structure of the adjacency matrix of the graph ($t=2$, $N_t=15$). 
Black regions are unit elements of the matrix.  
In white regions, all matrix elements are zeros. In grey regions, nonzero (unit) elements are present. The matrix is symmetric, and each column in grey blocks above the diagonal contains only two nonzero elements.
}
\label{f2}
\end{figure}

{\em Adjacency matrix}.---By definition, an element $a_{ij}$ of an adjacency matrix is equal to $1$ or $0$ depending on whether an edge between vertices $i$ and $j$ is present or not. The adjacency matrix $\hat{A}_t$ structure is schematically shown in Fig. \ref{f2}. At $t=-1$, this is the $2\times2$ matrix with zeros on the diagonal and two unit elements. At timestep $t$, we add rows and columns 
$i,j=N_{t-1}+1,\ldots,N_t$ (new vertices) to the matrix. Matrix is symmetric, and each unit element $a_{ij}$ above the diagonal of the matrix $\hat{A}_{t-1}$ generates, in addition, two unit elements $a_{is}$ and $a_{js}$ of $\hat{A}_t$. Here $N_{t-1}+1 \leq s \leq N_t$. Other elements are zeros. This produces the sparse block matrix shown in Fig. \ref{f2}. 

{\em Degree distribution}.---The degree spectrum of the graph is discrete: at time $t$, the number $m(k,t)$ of vertices of degree 
$k = 2, 2^2, 2^3, \ldots, 2^{t-1}, 2^t, 2^{t+1}$ is equal to 
$3^t, 3^{t-1}, 3^{t-2}, \ldots, 3^2, 3, 3$, respectively. 
Other values of degree are absent in the spectrum. Clearly, for the large network, $m(k,t)$ decreases as a power of $k$, so the network can be called ``scale-free''. Spaces between degrees of the spectrum grow with increasing $k$. 
Therefore, to relate the exponent of this discrete degree distribution to standard  
$\gamma$ exponent of a continuous degree distribution for random scale-free networks, we use a cumulative distribution 
$P_{cum}(k) \equiv \sum_{k^\prime \geq k} m(k^\prime,t)/N_t \sim k^{1-\gamma}$. Here $k$ and $k^\prime$ are points of the discrete degree spectrum. Thus we obtain 

\begin{equation}
\label{1}
\gamma = 1 + \frac{\ln3}{\ln2}
\, ,
\end{equation}  
so that $2<\gamma=2.585\ldots<3$. Compare $\gamma$ with the characteristic exponent in the relation between the ``mass'' and the ``perimeter'' of the graph. 
Also, notice that the maximal degree of a vertex is equal to 
$2^{t+1} \sim N_t^{\ln2/\ln3} = N_t^{1/(\gamma-1)}$, which coincides with a standard relation for the cutoff of degree distribution in growing scale-free networks \cite{dm01c}. 

{\em Distribution of clustering}.---By definition, the cluster coefficient $C$ of a vertex is the ratio of the total number of existing connections between all $k$ its nearest neighbors and the number $k(k-1)/2$ of all possible connections between them. Usually, only the average value of the clustering coefficient is considered. In our case, it is possible to obtain a more rich characteristic, namely, the distribution of the clustering coefficient in the graph. 

One can see that, in this graph, there is a one-to-one correspondence between clustering coefficient of a vertex and its degree: $C=2/k$. Thus, the number 
$m_c(C,t)$ of vertices with clustering coefficient $C = 1, 2^{-1}, 2^{-2}, \ldots, 2^{2-t}, 2^{1-t}, 2^{-t}$ is equal to $3^t, 3^{t-1}, 3^{t-2}, \ldots, 3^2, 3, 3$, respectively. In this case, it is natural to introduce the cumulative distribution of the clustering coefficient 
$W_{cum}(C) \equiv \sum_{C^\prime\leq C}  m_c(C^\prime,t)/N_t \sim C^{\ln3/\ln2} = C^{\gamma-1}$, where $C$ and $C^\prime$ are points of the discrete spectrum. This corresponds to power-law behavior of the corresponding continuous distribution of clustering $W(C) \sim C^{\,\gamma-2}$ for random scale-free network at small $C$. 

The average clustering coefficient can be easily obtained for arbitrary $t$,  

\begin{equation}
\label{2}
\overline{C}_t = \frac{4}{5}\,\frac{6^t + 3/2}{2^t (3^t + 1)}
\, .
\end{equation}  
For the infinite graph, 
$\overline{C} = 4/5$, so 
the clustering is high. 

{\em Degree correlations}.---The number $m(k,k^\prime,t)$ of edges, which connect vertices of degree $k$ and $k^\prime$, characterizes short-range degree-degree correlations in the graph. It is convenient to write $k\equiv2^{p+1}$ and use the notation $m(k,k^\prime,t)\equiv c(p,p^\prime,t)$. Then one can find directly:  

\begin{eqnarray}
\label{3}
& & 
c(t,t,t)=3 
\, , 
\nonumber
\\[3pt] 
& & 
c(t, p^\prime \leq t-1,t) = 3\cdot2^{t-1-p^\prime}
\, ,  
\nonumber
\\[3pt] 
& & 
c(p\leq t-1,p^\prime \leq p-1,t) = 3^{t-p}\, 2^{p-p^\prime-1} 
\, . 
\end{eqnarray} 
This yields the cumulative distribution $\sim k^{2-\gamma} k^{\prime\,-1}$ 
(we assume that $k \gg k^\prime$), which, in turn, corresponds to the effective continuous distribution 

\begin{equation}
\label{4}
P(k,k^\prime) \sim k^{1-\gamma}k^{\prime\,-2}
\, .
\end{equation} 
This expression coincides with the corresponding asymptotic formula 
for an arbitrary random scale-free citation graph \cite{dm01c} (by definition, a citation graph is a growing network, in which new edges do not emerge between pairs of old vertices). Originally, Eq. (\ref{4}) was obtained exactly for a specific model in Ref. \cite{kr00c}.

{\em Shortest-path length distribution}.---
Here we briefly outline our exact results 
for the distribution 
${\cal P}(\ell,t) \equiv n(\ell,t)/[ N_t (N_t-1)/2 ]$, where $n(\ell,t)$ is the number of pairs of vertices with minimal separation $\ell$. 
Details of the solution and general expressions for $n(\ell,t)$ will be published elsewhere.  
  
For calculation of $n(\ell,t)$ 
one may use the following property.  
The length $\ell_{ij}$ of the shortest path between vertices $i$ and $j$ is equal to the minimal power of the adjacency matrix with nonzero $\{ij\}$ element: $\{\hat{A}^{\ell-1}\}_{ij}=0, \ \{\hat{A}^\ell\}_{ij}\neq0$.  
This property allows us to obtain $n(\ell,t)$ by counting the total numbers of nonzero elements in sequential powers of the adjacency matrix. This yields $n(\ell,t)$ for several first steps: 

\begin{equation}
\label{5} 
\begin{array}{lllll}        
3                                                             
\\        
9    &    6 
\\       
27   &    57    &     21 
\\                                                   
81   &    351   &     369   &    60  
\\       
243  &    1806  &     3582  &    1716  &    156  
\\ 
\ldots &        &           &          &
\end{array}
\,  
\end{equation}  
where $t$ labels lines ($t=0,1,2,3,4,\ldots$) and $\ell=1,2,3,4,5,\ldots$ is the index of columns. 

The exact analytical form of the distribution ${\cal P}(\ell,t)$ was obtained by solution of recursion relations for $n(\ell,t)$. In particular, an exact expression for the average shortest-path length is of the form 

\begin{equation}
\label{5a}
\overline{\ell}(t \geq 0) = 
\frac{(4t+11)3^{2t} + 10\cdot 3^t + 3}{3(3^t + 1)(3\cdot 3^t + 1)}
\, .
\end{equation} 
One may check this expression using Eq. (\ref{5}). 
The distribution quickly approaches an asymptotic regime, where 

\begin{equation}
\label{5b}
\overline{\ell}(t \gg 1) = \frac{4}{9}\, t + \frac{11}{9} + {\cal O}(t\,3^{-t}) \cong 
\frac{4}{9}\ln N_t + \frac{5}{3}\ln\frac{3}{2}
\, .
\end{equation} 
Thus, the average shortest-path length logarithmically grows with increasing size of the graph. Expression (\ref{5b}) may be compared with the standard estimate \cite{nsw00}: $\overline{\ell} \sim \ln N /\ln\overline{k} = \ln N /\ln4$. The relative difference is surprisingly small, $(4/9)/(\ln4)^{-1}=0.616\ldots$. 
Notice that according to standard arguments \cite{ab01a,dm01c}, the classical formula is not applicable for $\gamma \leq 3$. 
Nevertheless, Eq. (\ref{5b}) demonstrates that the classical estimate is unexpectedly good in this case where degree-degree correlations are strong. 

At large $t$, the distribution takes the Gaussian form 

\begin{equation}
\label{5c}
{\cal P}(\ell,t) \cong 
\frac{1}{\sqrt{2\pi(2^2/3^3)t}} \exp
\left[ -\frac{(\ell - \overline{\ell}(t))^2}{2(2^2/3^3)t} \right]
\, ,
\end{equation} 
which is violated only in narrow regions of width $\sim t^{1/3}$ near the points 
$\ell=1$ and $\ell=t+1$. One sees that the width of the distribution 
is of the order of $\sqrt t \sim \sqrt{\ln N_t} \ll \overline{\ell}(t)$. 
Notice that the simulations of the Barab\'asi-Albert growing random network 
also yield  
a Gaussian-like ${\cal P}(\ell)$ \cite{k01}.

{\em Eigenvalue spectrum of the adjacency matrix}.---The spectrum 
$G(\lambda)$ contains $N_t$ 
eigenvalues. For $t\geq2$, $N_{t-1}-3$ 
of them are equal to zero and, for $t\geq3$, there are $N_{t-2}-3$ 
eigenvalues equal to $\sqrt2$ and the same number of those equal to $-\sqrt2$.
Here we do not derive analytical results for the entire eigenvalue spectrum but only study its tail using a simple numerical analysis. 

\begin{figure}
\epsfxsize=75mm
\centerline{\epsffile{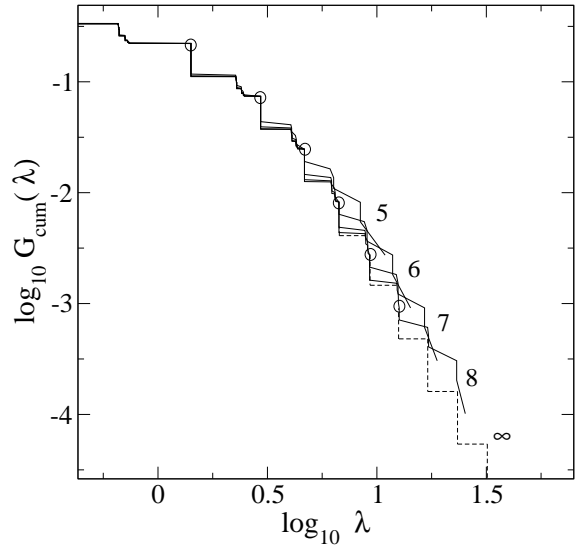}}
\caption{
Log-log plot of the cumulative distribution of eigenvalues of the adjacency matrix, 
$G_{cum}(\lambda) \equiv 
\sum_{\lambda^\prime\geq\lambda}G(\lambda^\prime)$. 
The curves show the spectra for $t=5,6,7,8$. The dashed line depicts the $t \to \infty$ limit. The $t$-independent points are marked.  
}
\label{f3}
\end{figure}

It is convenient to consider a cumulative distribution of eigenvalues 
$G_{cum}(\lambda) \equiv 
\sum_{\lambda^\prime\geq\lambda}G(\lambda^\prime)$. The results of numerical diagonalization of the adjacency matrix for several timesteps are shown in Fig. \ref{f3}. One sees that, in the large graph limit, the resulting cumulative distribution approaches a staircase-like form for $\lambda \gg 1$. 

We found network-size-independent points of the spectrum. Using the coordinates of these points, we calculated the series of slops of lines connecting these points in Fig. \ref{f3}: 
$1.49847, \ 2.38192, \ 3.03023, \ 3.40683, \ 3.53135, \ 3.557$ (the last value is actually a very good estimate obtained for $t=8$). Interpolation of these values yields the exponent $\delta-1=3.575\pm0.015$ of the cumulative distribution 
$G_{cum}(\lambda) \sim \lambda^{-(\delta-1)}$. This value is $2+\ln3/\ln2 = 1+\gamma = 3.585\ldots$ to within a precision of our numerics. The height of steps in Fig. \ref{f3} is $\log_{10}3$, their width approaches  
$[(2/\log_{10}3)+(1/\log_{10}2)]^{-1}$ for $\lambda \gg 1$. 

Thus, one can suggest that exponent $\delta$ of the corresponding continuous eigenvalue spectrum $G(\lambda \gg 1) \sim \lambda^{-\delta}$ is 
$\delta = 2+\gamma$. One should mention that the direct study of the eigenvalue spectrum for a growing random network with $\gamma=3$ (the Barab\'asi-Albert model) showed power-law dependence in a too narrow range of $\lambda$ to make precise conclusions \cite{fdbv01} (see also Ref. \cite{gkk01}). An estimate for the exponent in this situation was $\delta\approx5$ \cite{fdbv01}, which supports our conjecture.

{\em Percolation properties}.---Let us delete at random vertices or edges of the pseudofractal. One may check that, in the infinite network limit, the giant connected component of the graph disappears only if ``almost all'' vertices or edges are removed. 
In the language of mathematical graph theory this means that the fraction of 
vertices (edges), which we have to delete, approaches one as $N_t \to \infty$. 
This is a standard property of scale-free networks with $\gamma \leq 3$ \cite{ceah00a}.

{\em Discussion}.---The network that we study in this Letter is a citation graph. Each new edge connects a new vertex and an old one. This is actually a deterministic variation of the scale-free growing network \cite{dms003} in which one vertex is created per unit time and connects to both the ends of a randomly chosen edge. Therefore, it is not so strange that the properties of the pseudofractal network resembles those of scale-free random citation graphs. However, 
it is really surprising how close they appear to each other. Hardly one can propose more simple deterministic scale-free growing network. 
Therefore, we have found a very convenient tool for exploration of complex scale-free networks. 
    
The extreme simplicity of the pseudofractal graph has allowed us to obtain a number of new results for growing networks. In particular, for this network with strong correlations, we have obtained the exact shortest-path-length distribution and the eigenvalue spectrum of the complex adjacency matrix. From the latter, we have made a conjecture 
that exponent of eigenvalue spectra of scale-free citation graphs is 
$2+\gamma$. Actually, we have shown that randomness of scale-free growing networks is of secondary importance for their structure. 

S.N.D. thanks PRAXIS XXI (Portugal) for a research grant PRAXIS XXI/BCC/16418/98. S.N.D and J.F.F.M.  
were partially supported by the project POCTI/99/FIS/33141. We also thank A.N. Samukhin and A. Krzywicki for useful discussions. 
\\

\noindent
$^{\ast}$      E-mail address: sdorogov@fc.up.pt \\
$^{\dagger}$   E-mail address: goltsev@gav.ioffe.rssi.ru \\
$^{\ddagger}$  E-mail address: jfmendes@fc.up.pt

\end{multicols}

\end{document}